\documentclass[%
 reprint,
 amsmath,amssymb,
 aps,
prl,
]{revtex4-1}

\usepackage[hidelinks]{hyperref}
\usepackage{graphicx}
\usepackage{dcolumn}
\usepackage{bm}
\usepackage{mathtools}
\usepackage{comment}
\usepackage{overpic}
\DeclareMathOperator*{\argmax}{argmax}
\DeclareMathOperator*{\argmin}{argmin}

\newcommand{\inv}{\mathcal{I}}

\newcommand{\de}{\textrm{d}}
\newcommand{\Tr}{\textrm{Tr}}
\newcommand{\avg}[1]{\left\langle #1 \right\rangle}

\newcommand{\Rey}{\textrm{Re}}

\newcommand{\aMOD}{a}
\newcommand{\ADNS}{A}
\newcommand{\AMOD}{\mathcal{A}}

\newcommand{\fDNS}{f}

\begin{document}

\title{Tailor-designed models for the turbulent velocity gradient through normalizing flow}

\author{M.~Carbone${}^{1,2}$}
\email{maurizio.carbone@uni-bayreuth.de}
\author{V.~J.~Peterhans${}^{3,2}$}
\author{A.~S.~Ecker${}^{4,2}$}
\author{M.~Wilczek${}^{1,2}$}
\email{michael.wilczek@uni-bayreuth.de}
\affiliation{${}^{1}$Theoretical Physics I, University of Bayreuth, Germany\\
${}^{2}$Max Planck Institute for Dynamics and Self-Organization, G\"ottingen, Germany\\
${}^{3}$Faculty of Physics, University of G\"ottingen, Germany\\
${}^{4}$Institute of Computer Science and Campus Institute Data Science, University of G\"ottingen, Germany}

\date{\today}

\begin{abstract}
Small-scale turbulence can be comprehensively described in terms of velocity gradients, which makes them an appealing starting point for low-dimensional modeling. Typical models consist of stochastic equations based on closures for non-local pressure and viscous contributions. The fidelity of the resulting models depends on the accuracy of the underlying modeling assumptions. 
Here, we discuss an alternative data-driven approach leveraging machine learning to derive a velocity gradient model which captures its statistics by construction.
We use a normalizing flow to learn the velocity gradient probability density function (PDF) from direct numerical simulation (DNS) of incompressible turbulence. Then, by using the equation for the single-time PDF of the velocity gradient, we construct a deterministic, yet chaotic, dynamical system featuring the learned steady-state PDF by design. Finally, utilizing gauge terms for the velocity gradient single-time statistics, we optimize the time correlations as obtained from our model against the DNS data.
As a result, the model time realizations statistically closely resemble the time series from DNS.
\end{abstract}


\maketitle

Velocity gradients encode comprehensive information on the smallest scales of turbulent flows \cite{Meneveau2011,johnson2024arfm}, their statistics unveiling hallmarks of turbulence including cascades \cite{Alexakis2018,Carbone2020a,Johnson2020,johnson2024arfm}, time irreversibility \cite{Jucha2014,Cheminet2022} and intermittency \cite{She1990,Buaria2020,Fuchs2022}.
On a more applied level, velocity gradients also govern the small-scale turbulent advection, stretching, and folding of material elements \cite{girimaji1990b,Bentkamp2022,Qi2023}, and the collisional statistics of inertial particles transported by the flow \cite{Balkovsky2001,Bec2007,Ireland2016}, relevant e.g.~for droplet growth in turbulent warm clouds \cite{Bodenschatz2010,Grabowski2013}.
Reduced-order modeling of the velocity gradient is thus appealing for a better theoretical understanding of turbulence and for practical purposes.

A widely-employed and successful approach consists of modeling the Lagrangian dynamics of the velocity gradient $\bm{A}\equiv \bm{\nabla u}$ (where $\bm{u}(\bm{x},t)$ is a three-dimensional turbulent velocity field), following fluid particle trajectories \cite[e.g.][]{Girimaji1990a,Chertkov1999,Naso2005Scale,Chevillard2006,Wilczek2014,Johnson2016,Johnson2017Turbulence,Leppin2020}.
However, this single-time/single-point modeling strategy has to deal with the closure problem:
Due to the intrinsic non-locality of turbulence, the dynamical evolution of the velocity gradient at a point is not uniquely determined solely by the velocity gradient configuration at that point.
This issue is evident from the equations governing the velocity gradient dynamics
\begin{align}
\de_t \bm{A} = -\bm{A}^2 - \bm{H} + \nu\nabla^2 \bm{A} + \bm{\nabla F}, && \Tr(\bm{A}) &= 0,
\label{eq_grad}
\end{align}
where $\de_t$ indicates the Lagrangian derivative taken following the fluid particle trajectory, $\bm{\nabla F}$ represents an external forcing, $\Tr$ indicates the matrix trace and standard matrix product is implied. The pressure Hessian $\bm{H}$ and the viscous Laplacian $\nu\nabla^2 \bm{A}$ are nonlocal functionals of the velocity gradient field \cite[e.g.][]{ohkitani1995,Majda2001}. These unclosed terms cannot be computed by knowing the local velocity gradient only, and they require modeling when addressed from a single-time/single-point perspective.

Due to the intrinsic randomness of equations \eqref{eq_grad}, reduced-order modeling of the velocity gradient aims to predict the statistics of $\bm{A}(t)$, rather than the individual time realizations. This is achieved by integrating a model Langevin equation \cite{Girimaji1990a,Chertkov1999,Chevillard2006,Wilczek2014,Johnson2016,Leppin2020}
\begin{align}
    \de_t\bm{\AMOD}  = \bm{N}\!\left(\bm{\AMOD}\right) + \bm{\Gamma},
    && \Tr(\bm{\AMOD})=0,
    \label{eq_mod}
\end{align}
for an ensemble of initial conditions, 
where $\bm{N}$ contains a closure model for the pressure Hessian and viscous term, and $\bm{\Gamma}$ is a Gaussian noise, such that the steady-state statistics of the $3 \times 3$ matrix $\bm{\AMOD}$ approximate the statistics of the turbulent velocity gradient ensemble $\bm{\ADNS}$.
The modeling step from Eq.~\eqref{eq_grad} to \eqref{eq_mod} implies that the right-hand side of \eqref{eq_grad} depends on the whole velocity field $\bm{u}(\bm{x},t)$, while the right-hand side of \eqref{eq_mod} depends only on the eight-dimensional state variable $\bm{\AMOD}$. This drastic reduction of degrees of freedom generates the aforementioned closure problem.

Many phenomenological models have been successful in qualitatively predicting the statistical dynamics of the velocity gradient in isotropic turbulence \cite[e.g.][]{Girimaji1990a,Chertkov1999,Naso2005Scale,Chevillard2006,Wilczek2014,Johnson2016,Johnson2017Turbulence,Pereira2018,Leppin2020,Johnson2018Predicting}. More recently, data-driven approaches based on machine learning \cite{Tian2021,Buaria2023} or lookup tables \cite{Das2023} have reached accuracy levels comparable to the most advanced phenomenological closures.
These recent approaches for data-driven models \cite{Tian2021,Buaria2023,Das2023} approximate/learn the unclosed terms arising from \eqref{eq_grad}. Statistically, this amounts to estimating the average time derivative of the velocity gradient conditional on the configuration of the velocity gradient itself, i.e.~$\avg{\de_t \bm{A}(t)|\bm{A}}$, and then integrating it in time to generate velocity gradient time realizations. A major issue in this procedure is that an accurate a-priori estimation of the conditional time derivatives does not guarantee that the a-posteriori generated velocity gradient time series are realistic. This a-priori vs.~a-posteriori issue, also known from Large Eddy Simulations \cite{Jimenez2000}, leads not only to inaccuracies in the predictions but may even imply the blow-up of the gradient realizations over time.
Diverging trajectories are either neglected \cite{Buaria2023} or damped via ad-hoc terms \cite{Leppin2020} to reach a non-trivial statistically steady state.
Another open problem of the current phenomenological and data-driven models is that they can faithfully approximate single-time statistics, performing well on certain marginal PDFs of the relatively high-dimensional velocity gradient PDF, while leaving room for improvement on other marginal distributions and two-time statistics (e.g.~\cite{Chevillard2006,Wilczek2014,Johnson2016,Buaria2023}).
On a more general level, recent alternatives that have the potential of improving the classical reduced-order modeling include Lagrangian Large Eddy Simulations \cite{Tian2023} and time series generation by diffusive models \cite{Ho2020,Li2023} or Generative Adversarial Networks \cite{Goodfellow2014,Arjovsky2017}.

In this work, we propose an approach to construct velocity gradient models that yield the correct velocity gradient statistics of homogeneous isotropic turbulence by design. Earlier attempts to formulate such constrained models consisted of imposing certain statistics such as the log-normality of the velocity gradient magnitude \cite{Girimaji1990a,Pereira2018}.
Instead, here we constrain the model to yield the full distribution of the velocity gradient: We train a neural network to predict the PDF of the velocity gradient, $f(\bm{\ADNS})$, from a DNS dataset employing a normalizing flow \cite{Tabak2010,Kobyzev2021}. Then we construct a dynamical system in the form of Eq.~\eqref{eq_mod} that has the learned PDF as the steady-state probability density distribution.
In the following, we focus on a deterministic reduced-order model capable of maintaining a chaotic statistically steady state (i.e.~Eq.~\eqref{eq_mod} without noise). This eliminates the need for any ad-hoc stochastic forcing, which is required in existing models to keep non-trivial steady-state statistics \cite[e.g.][]{Chevillard2006,Wilczek2014,Johnson2016,Leppin2020,Tian2021,Das2023,Buaria2023}.
Our deterministic model is associated with a Liouville equation for the conservation of probability in the phase space
\begin{align}
\partial_t\fDNS + \partial_{ij} \left(N_{ij} \fDNS\right) = 0,
\label{eq_FPE}
\end{align}
where $\partial_{ij}$ denotes tensor derivative with respect to $\AMOD_{ij}$, and summation over repeated indices is implied.
Since we know the intended steady-state solution $\fDNS(\bm{\AMOD})$, we can derive an expression for the vector field $\bm{N}$ such that Eq.~\eqref{eq_FPE} yields that solution by construction, namely
\begin{align}
N_{ij}(\bm{\AMOD};\{\psi\}) &= \partial_{pq} T_{ijpq}  + T_{ijpq}\partial_{pq} \log \fDNS,
\label{eq_vect_field}
\end{align}
where $\bm{T}(\bm{\AMOD};\{\psi\})$ is any smooth tensor function of the velocity gradient such that $T_{ijpq} = -T_{pqij}$, and $\{\psi\}$ is a set of model parameters.
For stochastic generalizations of our approach, the same can be done by replacing $\bm{T}$ with $\bm{T}+\bm{D}/2$ in Eq.~\eqref{eq_vect_field}, where $\bm{D}(\bm{\AMOD})$ is the correlation of the noise $\bm{\Gamma}$ in Eq.~\eqref{eq_mod}.

A drift of the form of Eq.~\eqref{eq_vect_field} ensures that $\fDNS$ is a steady-state distribution of our model, and this holds true for any tensor $\bm{T}$ that is anti-symmetric in its index pairs. Therefore, imposing the single-time steady PDF leaves the gauge term $\bm{T}$ undetermined, that is, there exists a class of models with the same single-time distribution but different multi-time statistics. Within this class, we can select the model that produces realizations of the velocity gradient which are statistically closest to the realizations in the DNS dataset. To this end, we optimize the conditional time derivatives together with the time correlations of the velocity gradient.
For statistically isotropic turbulence, we can exploit additional gauge terms that affect velocity gradient multi-time correlations while leaving the single-time statistics unchanged for kinematic reasons \cite{Carbone2020b,Leppin2020}.
Those gauge terms stem from the particular functional form of an isotropic velocity gradient PDF, which depends only on the five invariants formed through the strain rate $\bm{S}=(\bm{A}+\bm{A}^\top)/2$ and the vorticity $\bm{\omega}=\nabla\times \bm{u}$
\cite{Pennisi1987,Lund1992}
\begin{align}
\inv_1 &= \Tr\big(\bm{S}^2 \big) &&
\inv_2 =  \|\bm{\omega}\|^2/2 &&
\inv_3 = \Tr\big(\bm{S}^3 \big) &&
\nonumber\\
\inv_4 &= \bm \omega \cdot \bm{S \omega}
&&
\inv_5 = \bm{\omega} \cdot \bm{S}^2\bm{\omega}.
\label{eq_inv}
\end{align}
The tensors $\widetilde{\bm{T}}$ which leave the dynamics of the invariants \eqref{eq_inv} unchanged satisfy the kinematic condition
\begin{align}
\widetilde{T}_{ijpq}\partial_{pq} \inv_k = 0,
\label{eq_gauge_T}
\end{align}
and will help us optimize the velocity gradient multi-time correlations while preserving its single-time statistics.

\begin{figure*}
    \begin{overpic}[width=1\linewidth]{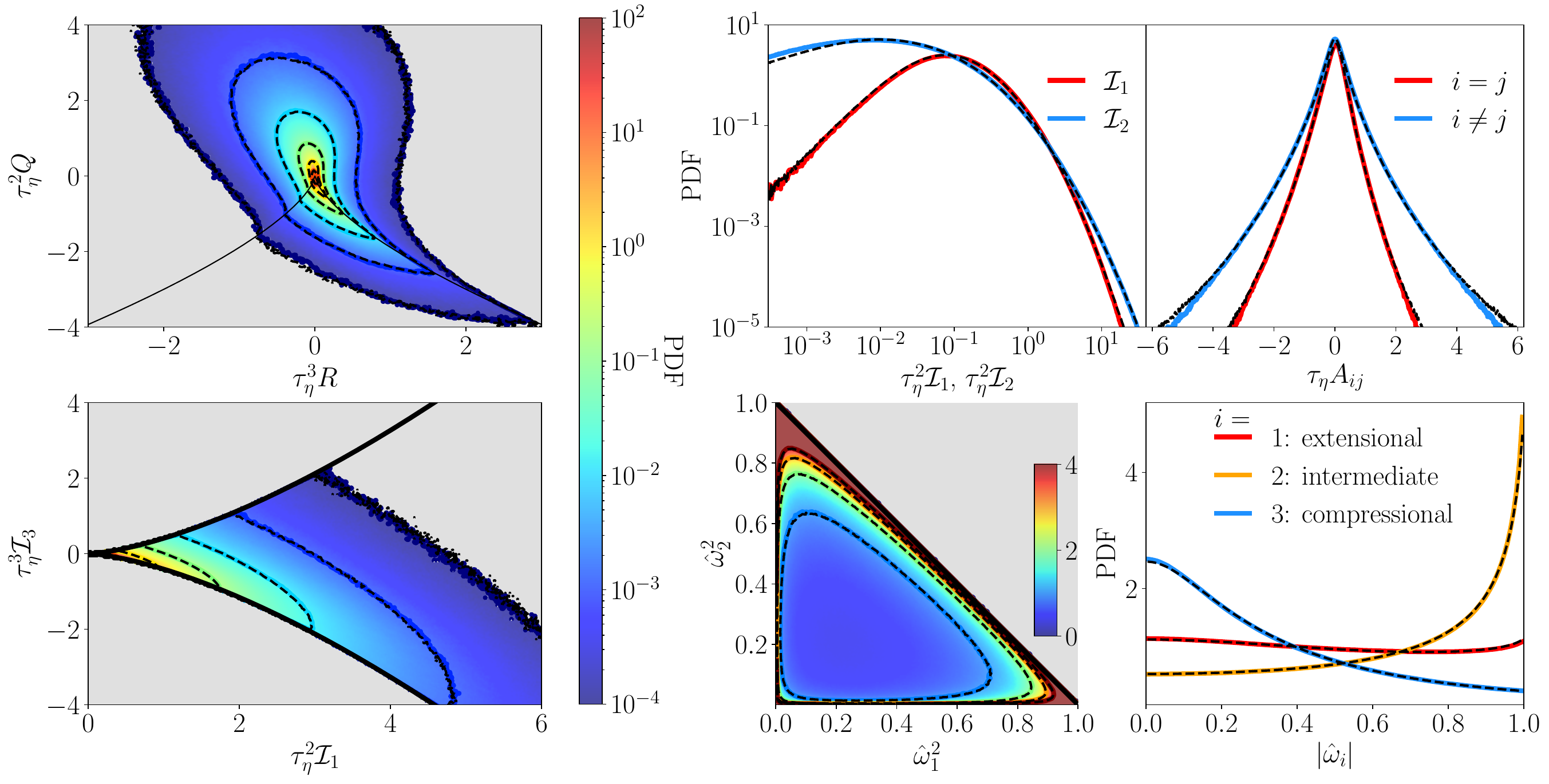}
    \put(32,47){(a)}
    \put(32,23){(b)}
    \put(68,47){(c)}
    \put(67,23){(e)}
    \put(93,47){(d)}
    \put(94,23){(f)}
    \end{overpic}
    \caption{Single-time velocity gradient distributions from the DNS (colormaps and colored solid lines) and from our model (black dashed lines). Panel (a) shows the joint PDF of principal invariants of the velocity gradient, $Q=-\Tr(\bm{A}^2)/2$ and $R=-\Tr(\bm{A}^3)/3$, panel (b) shows the joint PDF of the strain invariants $\inv_1 = \Tr\big(\bm{S}^2 \big)$ and $\inv_3 = \Tr\big(\bm{S}^3 \big)$.
    Panel (c) shows the PDF of the strain-rate and vorticity magnitudes; the PDFs of the longitudinal and transverse velocity gradient components are shown in panel (d). Panels (e) and (f) display the PDFs of the normalized vorticity components in the strain eigenframe, $\hat{\omega}_i=\bm{v}_i \cdot \bm{\omega}/\Vert\bm{\omega}\Vert$, where $\bm{v}_i$ are the strain-rate eigenvectors (index 1 denoting the most extensional direction and 3 the most compressional). All dimensional quantities are nondimensionalized by the Kolmogorov time scale, $\tau_\eta=1/\sqrt{\avg{2\inv_1}}$.}
    \label{fig_single_point}
\end{figure*}

Now that we have established the basic theory behind the method, let us discuss its practical implementation.
The proposed approach relies on a parameterization of the velocity gradient PDF, a technical step that we accomplish through a normalizing flow \cite{Tabak2010,Kobyzev2021}. 
To facilitate the implementation of the normalizing flow, we unroll the components of the $3 \times 3$ traceless matrix $\AMOD_{ij}$ into an eight-dimensional vector $\bm \aMOD$.
Then, starting from an ensemble of eight-dimensional Gaussian vectors $\bm{\aMOD}^{(0)}$,
we perform a sequence of real, non-volume-preserving, quasi-linear transformations  \cite{Dihn2017}.
In each transformation layer, we update one variable through an affine transformation, whose scaling $\Lambda^{(K)}$ and bias $b^{(K)}$ are computed from the remaining seven variables. Then we repeat the sequence of transformations four times for each variable
\begin{align}
\aMOD^{(K)}_i = 
\begin{cases}
\exp\left(\Lambda^{(K)}\right) \aMOD^{(K-1)}_i + b^{(K)} \textrm{ if } i = K \, (\bmod \, 8) \\
\aMOD^{(K-1)}_i \textrm{ otherwise},
\end{cases}
\label{eq_transf_layers}
\end{align}
from transformation layer $K=1$ to $K=32$.
The scaling functions
$\Lambda^{(K)}\left(\bm{\aMOD}^{(K-1)};\{\theta^{(K)}\} \right)$
and the bias functions
$b^{(K)}\left( \bm{\aMOD}^{(K-1)};\{\theta^{(K)}\} \right)$ in Eq.~\eqref{eq_transf_layers}
depend on all the gradient components except the one updated across that layer,
so that $\partial a_i^{(K)}/
\partial a_j^{(K-1)}=\exp(\Lambda^{(K)})$ for $i=K \, (\bmod \, 8)$ and $i=j$.
The scaling and bias functions consist of two neural networks for each layer, featuring two hidden linear layers with 64 neurons, Softplus activation function and parameters $\{\theta^{(K)}\}$.
The transformed eight-dimensional vector $\bm{\aMOD}^{(32)}$ is mapped back to a $3 \times 3$ traceless matrix, $\bm{\aMOD}^{(32)} \Leftrightarrow \bm{\AMOD}$, and the ensemble of matrices $\bm{\AMOD}$ aims to be statistically similar to the ensemble of turbulent velocity gradients $\bm{\ADNS}$, through a maximum likelihood criterion.
Owing the quasi-linearity of the transformation \eqref{eq_transf_layers}, the log-likelihood $\log \fDNS^{(K)}$ of the ensemble of matrices $\bm{\AMOD}^{(K)}$ changes across a layer as
\begin{align}
\log \fDNS^{(K)}\big(\bm \AMOD^{(K)}\big) = \log \fDNS^{(K-1)}\big(\bm \AMOD^{(K-1)}\big) + \Lambda^{(K)}.
\label{eq_transf_PDF}
\end{align}
We find the optimal parameters in the transformations \eqref{eq_transf_layers} and the corresponding parameterization of the velocity gradient PDF by maximizing the log-likelihood of the turbulent velocity gradient ensemble from the DNS
\begin{align}
\{\bar{\theta}\} = \argmax_{\{\theta\}}
\avg{\log \fDNS\left(\bm{\ADNS};\{\theta\}\right)},
\label{eq_max_like}
\end{align}
where  $\fDNS \equiv \fDNS^{(32)}$ is the target velocity gradient PDF, $\bm{A}$ is an element of the turbulent velocity gradient ensemble from the DNS, the angle brackets denote ensemble average computed over the DNS dataset, and $\{\theta\}$ stands for all the parameters featured in the transformations \eqref{eq_transf_layers}.

The DNS dataset consists of the velocity gradient realizations along $2^{18}$ fluid particle Lagrangian trajectories from a DNS of three-dimensional, incompressible, statistically steady and isotropic turbulence. The velocity gradient time realizations span $64\tau_\eta$, and are sampled with a time step $\tau_\eta/60$ (where $\tau_\eta$ is the Kolmogorov time scale).
The DNS resolves $1024^3$ Fourier modes with a maximum resolved wavenumber $k_{\textrm{max}}=3/\eta$ (where $\eta$ is the Kolmogorov scale) and Reynolds number based on the Taylor microscale $\Rey_\lambda=220$.
The normalizing flow implementation relies on Pytorch \cite{Paszke2019pytorch}. We employ the Adam optimizer \cite{kingma2017adam} with a learning rate of $5\times 10^{-5}$ and running average coefficients $\beta_1=0.9$, $\beta_2=0.999$. We train the model on 40 GPUs for $5\times 10^4$ training steps, on batches of 1024 velocity gradient time realizations.

\begin{figure*}
    \begin{overpic}[width=1\linewidth]{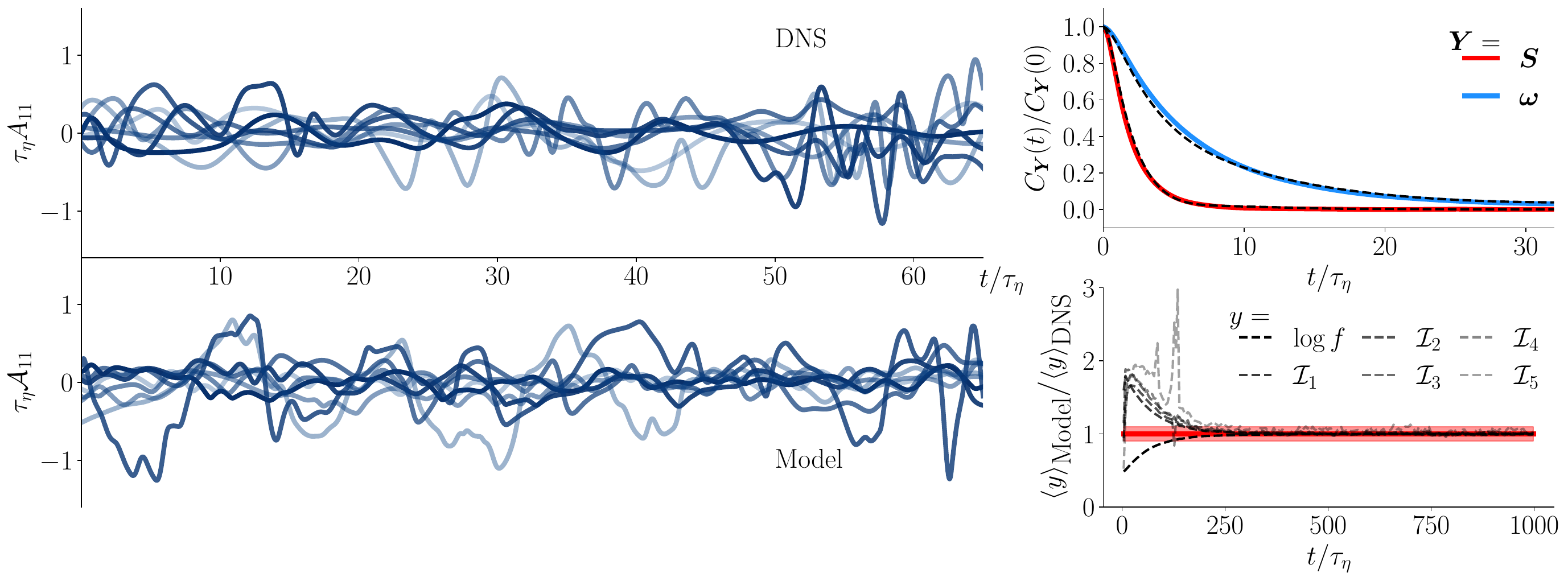}
    \put(46,34.1){(a)}
    \put(46,7.2){(b)}
    \put(72,35){(c)}
    \put(72,18){(d)}
    \end{overpic}
    \caption{Comparison of DNS and model trajectories, time correlations and model convergence. Panels (a) and (b) show sample realizations of the longitudinal velocity gradient component along fluid particle trajectories from the DNS and our model. Panel (c) displays the normalized time correlations of the strain rate and vorticity from the DNS (colored solid lines) and our model (black dashed lines). Panel (d) shows the convergence of our model (transparent black dashed lines) from Gaussian initial conditions to turbulent-like statistics, as quantified by the log-likelihood of the velocity gradient ensemble and the average velocity gradient invariants, as functions of time. The red line indicates the DNS reference, with the transparent region denoting a 10\% accuracy interval. All dimensional quantities are nondimensionalized by the Kolmogorov time scale, $\tau_\eta=1/\sqrt{\avg{2\inv_1}}$.}
    \label{fig_trajectories}
\end{figure*}
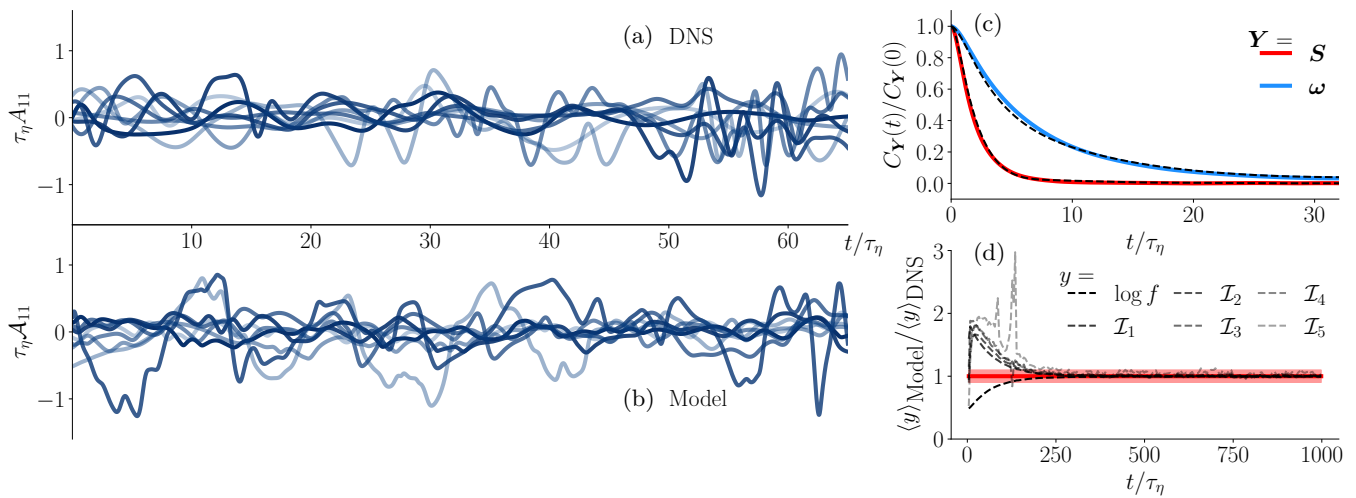

The normalizing flow yields an accurate parameterization of the turbulent velocity gradient PDF, as well as an invertible transformation \eqref{eq_transf_layers} capable of mapping an ensemble of Gaussian random matrices onto an ensemble resembling the turbulent velocity gradients. This is in the same spirit of Gaussian mapping closures in fluid dynamics \cite{Chen1989,Kraichnan1990}.
We can now sample the learned steady-state PDF by using the transformation \eqref{eq_transf_layers} or by integrating our model \eqref{eq_mod} with vector field \eqref{eq_vect_field}, and the results coincide. 
Figure \ref{fig_single_point} shows a comparison between velocity gradient statistics from DNS data and our model. The model captures several hallmark statistics of small-scale turbulence with unprecedented accuracy, including:
The PDF of the velocity gradient (Fig.~\ref{fig_single_point}a) and strain-rate principal invariants (Fig.~\ref{fig_single_point}b),
the distribution of the strain-rate and vorticity magnitude (Fig.~\ref{fig_single_point}c) and the velocity gradient Cartesian components  (Fig.~\ref{fig_single_point}d),
as well as the statistical alignments between the vorticity and the strain-rate eigenframe (Fig.~\ref{fig_single_point}e-f). 
The quantitative agreement between the model and the DNS data on several marginal distributions from the velocity gradient PDF shows that we have gained access to the full velocity gradient PDF.

Our data-driven approach decouples the single-time statistics modeling from the optimization of the multi-time statistics.
The anti-symmetric tensor $\bm{T}(\bm{\AMOD};\{\psi\})$ governs the velocity gradient multi-time statistics, and we represent it by a neural network featuring three hidden linear layers with 128 neurons, Softplus activation function and parameters $\{\psi\}$.
To optimize the parameters of $\bm{T}$, we numerically integrate the model equations \eqref{eq_mod} for an ensemble of initial conditions sampled from the DNS data, $\bm{\AMOD}(0)\equiv \bm{\ADNS}(0)$, and require that the time realizations $\bm{\AMOD}(t)= \bm{\AMOD}(0) + \int_0^t\de s \, \bm{N}(\bm{\AMOD}(s);\{\psi\})$ have time correlations and conditional time derivatives similar to the DNS,
\begin{align}
&\{\bar{\psi}\} = \argmin_{\{\psi\}}
\Bigg(
\int_0^T\de t
\left\Vert 
\avg{\AMOD_{ij}(0)\AMOD_{pq}(t) 
-
\ADNS_{ij}(0)\ADNS_{pq}(t)}
\right\Vert^2
\nonumber\\
&+\beta\sum_{k=1}^5\avg{\left\Vert
\partial_{ij} \inv_k \left(\bm{\ADNS}\right)
\left(
\de_t\ADNS_{ij}-N_{ij}\left(\bm{\ADNS};\{\psi\}\right)
\right)
\right\Vert^2}
\Bigg)
\label{eq_t_opt}
\end{align}
where $\beta$ is a constant weighting parameter.
For the minimization \eqref{eq_t_opt}, we employ the Adam optimizer with the parameters reported above for Eq.~\eqref{eq_max_like}, a time span $T=8\tau_\eta$ and weighting factor $\beta=10^{-5}$. The training converges within $10^{3}$ train steps.

The optimization \eqref{eq_t_opt} results in a network $\bm{T}$, thus a tensor function $\bm{N}$ (see Eq.~\eqref{eq_vect_field}), such that the time correlations from the model 
approximate the time correlations from the DNS data while preserving the previously learned single-time PDF $\fDNS(\bm{\AMOD})$ by construction.
Integrating the model \eqref{eq_mod} with the drift \eqref{eq_vect_field} optimized according to the criterion \eqref{eq_t_opt} can be understood as a particular Monte Carlo sampling of the PDF $\fDNS$ aided by normalizing flow \cite{Bialas2023QCD}, in which the sampling trajectories $\bm{\AMOD}(t)$ have the meaning of actual time realizations.
Although the model was not designed to capture individual velocity gradient time realizations, the time series  generated by the model look qualitatively similar to those from DNS, as displayed in Fig.~\ref{fig_trajectories}(a-b).
By visual inspection, the DNS trajectories $\ADNS_{11}(t)$ and model trajectories $\AMOD_{11}(t)$ have similar amplitudes with large and negative excursions occurring more frequently than positive ones, and the duration of such excursions is comparable.
Also quantitatively, the autocorrelation functions of strain $C_{\bm S} = \avg{S_{ij}(0)S_{ij}(t)}$ and vorticity $C_{\bm \omega} = \avg{\omega_{i}(0)\omega_{i}(t)}$ from the model match well those of the DNS data, as Fig.~\ref{fig_trajectories}(c) shows.

Since our model is deterministic, it is a priori unclear whether it approaches a unique statistically steady state independent of initial conditions. To answer this question, we performed numerical tests in which we observed convergence from an ensemble of Gaussian initial conditions to the prescribed steady-state PDF.
We quantify the convergence by comparing the log-likelihood of the model PDF $\avg{\log f(\bm{\AMOD}(t))}$ to the log-likelihood of the DNS data estimated via normalizing flow. Additionally, we compare the average invariants from the model realizations and the DNS data. As Fig.~\ref{fig_trajectories}(d) shows, all quantities approach the expected steady-state values for times longer than ${\cal O}(100 \tau_\eta)$. The convergence of the deterministic system to a statistically steady state, independent of the initial conditions, is a typical feature of a chaotic system~\cite{Strogatz1994}. Accordingly, we find the largest Lyapunov exponent for our model to be positive, and the dynamics to be mildly dissipative, with an estimated Kaplan-Yorke dimension of $7.96$. Those observations, combined with the  convergence to stationary statistics, indicate deterministic chaos.

Finally, the convergence to a target distribution showcases that it is possible to construct generative machine-learning models which do not rely on mapping stochastic signals into a distribution of target objects \cite[e.g.][]{Ho2020}, but rather use the chaotic nature of the underlying high-dimensional mapping to mimic creativity.

Summarizing, we have developed a reduced-order deterministic velocity gradient model that captures the statistical properties of the full eight-dimensional velocity gradient PDF by design. This is achieved in a two-step procedure. First, the full PDF is learned from DNS data using a normalizing flow \cite{Tabak2010,Dihn2017}. Based on that, we construct a Liouville equation which has the learned PDF as a stationary solution. The Liouville equation is associated with a deterministic reduced-order model, which allows to generate time realizations of the process. We then use gauge freedom for the single-time statistics to optimize the time correlations and conditional time derivatives of the model to emulate actual realizations of Lagrangian velocity gradients in turbulence. The resulting deterministic reduced-order model shows signs of chaotic dynamics. For example, it features a positive Lyapunov exponent and converges to the expected statistically stationary state from Gaussian initial conditions. A more detailed analysis of the model dynamical system's properties is an interesting subject for future work.

\begin{acknowledgments}
This project has received funding from the European Research Council (ERC) under the European Union’s Horizon 2020 research and innovation programme (Grant agreement No.~101001081).
The authors gratefully acknowledge the scientific support and HPC resources provided by the Erlangen National High Performance Computing Center (NHR@FAU) of the Friedrich-Alexander-Universität Erlangen-Nürnberg (FAU) under the NHR project b159cb EnSimTurb.
Computational resources from the Max Planck Computing and Data Facility and support by the Max Planck Society are gratefully acknowledged.
\end{acknowledgments}

\bibliography{bibliography_ML}

\end{document}